\documentclass[a4paper]{jpconf}
\usepackage{graphicx}
\begin{document}
\title{Probing the Accretion Induced Collapse
of White Dwarfs in Millisecond Pulsars}

\author{Ali Taani}
\address{Applied Science Department, Aqaba University College, Al-Balqa Applied
University, Aqaba, Jordan}
\ead{ali.taani@bau.edu.jo}
\author{Awni Khasawneh}

\address{Royal Jordanian Geographic Center, Amman, 11941 Jordan}

\begin{abstract}

This paper investigates the progenitors of Millisecond Pulsars (MSPs) with a distribution of long
orbital periods (P$\rm_{orb} > 2$ d), to show the link between white
dwarf (WD) binaries and long
orbits for some binary MSPs through the Accretion Induced Collapse (AIC) of a WD. For this purpose, a model is presented %we present a model that attempts
to turn binary MSPs into wide
binaries and highly circular orbits (e $<$ 0.1) through the asymmetric kick imparted to the
pulsar during the AIC process, which may indicate a sizeable kick velocity along the
rotation of the proto-neutron star. The results show the effects of shock wave, binding energy, and mass loss (0.2M$_{\odot}$).
The model shows the pulsar systems are relevant to AIC-candidates.
%If the system survives, the resulting system typically has long circular orbit orbit  ($e < $ 0.1). Or it may disrupt the system and create a single MSP.
\end{abstract}

\section{Introduction}

 A number of channels of Millisecond Pulsars (MSPs) formation have been discussed $^{[1-3]}$.
The Recycling process is usually considered as a standard model to produce MSPs. An old and quiescent Neutron Stars (NSs) is spun up to
millisecond periods, through the accretion of matter from a
companion when the system is in the low-mass X-ray binary (LMXB)
phase of its evolution $^{[4-5]}$.
 Another often discussed theory involves the Accretion
Induced Collapse (AIC) of an O-Ne-Mg white dwarf (WD) in binaries $^{[6-10]}$, which can indicate that (1) the WD is massive enough to be
an O-Ne-Mg one, and thus in principle may be able to collapse to an
NS (e.g. Cal 83 $\rm \sim 1.3M_{\odot}$ and RX J0648.0–4418 $\rm
\sim 1.28M_{\odot}$), and (2) the mass transfer rate is large enough
to make the WD grow -by considering the critical value ($\rm \Delta
M_{crit} \sim 0.1M_{\odot}- 0.2M_{\odot}$)$^{[11]}$, because at low accretion rates Nova explosions and a WD will not grow.
But to make a Type Ia SNe (a thermonuclear explosion of a CO WD that has grown to the Chandrasekhar limit), the WD needs to grow, so it needs a high accretion
rate, such that steady nuclear burning on the surface ensues and the WD grows$^{[12]}$.% A characteristic example of this process is the
%first-discovered binary MSP PSR 1953+29 $^{[12]}$, %with P$\rm _{orb}$ = 117.35 d and B $\sim
%10^{8}$ G. The NS was most probably formed by the AIC of %a WD.
The time lines for the two routes of
MSP evolutions, recycling and AIC, are depicted in Fig. 1.

\begin{figure}
\includegraphics[width=12.0cm, angle=0] {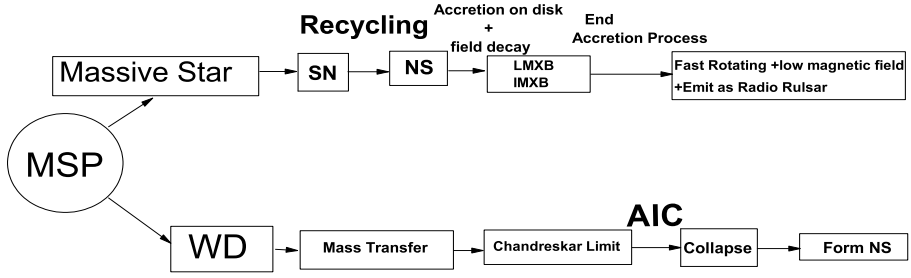}
\caption{The flow-chart illustrates various evolutionary scenarios
for MSPs involving recycling and AIC. }\label{taani}
\end{figure}

\section{Estimation of the new orbital period}

The relation between the orbital eccentricity and the amount of mass
$\rm \Delta M_{SNe}$ ejected in the SNe is
\begin{equation}
e= \frac{\Delta M_{SNe}}{M_{1}+M_{2}}
\end{equation}
where $M_{1}$ and $M_{1}$ are the masses of the first- and the
second-born NSs.

According to the energy conservation law$^{[13]}$

\begin{equation}
  \frac{GM}{R_{WD}}-\frac{GM}{L}=\frac{1}{2}[v_{final}^{2}-v_{initial}^{2}]
\end{equation}
where $G$ is the gravitational constant, $M$ the mass companion,
$R_{WD}$ the WD radius, $L$ the distance between the companions,
$v_{initial}$ the initial explosion velocity for the WD (it has been supposed that the
$v_{initial}\sim 0.1c$$^{[14]}$), and $v_{final}$ the velocity when the companion received the material on the surface; because the SNe
ejecta may either directly strip material from the companion by
direct transfer of momentum or evaporate the envelope through the
conversion of the blast kinetic energy into internal heat$^{[15]}$.
It has been assumed that the amount of ejected mass during the explosion
process by a WD is $\rm 0.1M_{\odot}$.

\begin{equation}
  \Delta M_{SNe}=\frac{R^{2}}{L^{2}}\times0.1M_{\odot}
\end{equation}

During the SNe, the WD will eject matter in every
direction, so $\Delta M_{SNe}$ is the amount of mass received by the companion (accreted material) and
$R$ is the radius of the companion. When $L\gg
R_{WD}$, after some calculations, Eq. 2 will be
\begin{equation}
\frac{R_{S}~c^{2}}{2R_{WD}}=10^{-4}c^{2},
\end{equation}
where $R_{S}$ is Schwartzchild radius.
%$\frac{GM}{R_{WD}}=\frac{1}{2}[v_{final}^{2}-v_{initial}^{2}]$,
%$~~\sim\frac{R_{S}~c^{2}}{2R_{WD}}=10^{-4}c^{2},$ where $R_{S}$ is
%Schwartz radius ($R_{S}=\frac{2GM}{c^{2}}\sim3$ km for
%($M/1M_{\odot}$)

It has been assumed that
the explosion to be instantaneous (short duration) like in very close binaries$^{[13]}$, in the range of LMXBs,  Cataclysmic Variable-type
binaries (CVs) and
close double pulsars, such as the PSR 1913+16, J0737- 3039AB and the
close WD-pulsar system PSR 0655+64$^{[12]}$. Hence
$v_{final}\approx v_{initial}\approx0.1c$
\begin{eqnarray}
 v_{k}=\frac{\Delta Mv_{final}}{M}=\frac{R^{2}}{L^{2}}(M/0.1M_{\odot})^{-1}\cdot
 v_{final}
\end{eqnarray}

%by the loss of angular momentum by gravitational radiation, ultimately leading
%This estimate gives the specific %angular momentum required in the outer parts of the progenitor's core.
where $v_{k}$ is the kick velocity of the companion. It should be
noticed that if the angular momentum of the accreted material equals
the orbital angular momentum, due to the instability in the accretion
shock around a proto-NS. This would enhance the
 possibility for the companion to be kicked away$^{[16]}$. An assumption here is that a more
energetic explosion would be able to impart a greater initial
angular velocity onto the proto-NS than a lesser explosion, and the
result is typically an eccentric orbit with an orbital period of a
days. The escape velocity for the companion in the gravitational field of
the WD is
\begin{equation}\label{a}
 v=\sqrt{\frac{2GM_{WD}}{L}}\simeq\sqrt{\frac{R_{S}}{L}}~~c
\end{equation}

Hence, this is the largest kick velocity that a system can attain
after a symmetric SNe explosion in case of $v_{k} \gg v$. Now let us define the $\eta$ parameter as the ratio between two
velocities
\begin{eqnarray}\label{b}
\eta=\frac{v}{v_{k}}\sim10^{-2}(\frac{R}{L})^{2}(\frac{L}{R_{S}})^{1/2}(\frac{M}{M_{\odot}})^{-1}
.
\end{eqnarray}
Assuming canonical values for $L\sim3\times10^{10}$ cm,
$R\sim10^{9}$ cm and $R_{S}=3\times10^{5}$ cm,
$\eta\sim(\frac{M}{0.1M_{\odot}})^{-1}=1$. The energy difference between the initial and final positions of the
companion in the explosion is
\begin{eqnarray}\label{difference}
\frac{GM_{WD}}{2L}-\frac{GM_{WD}}{2L_{1}}=\frac{1}{2}v^{2} .
\end{eqnarray}
Then by substituting $R_{S}$, Eq.~(\ref{a}) and ~(\ref{b}) into
Eq.~(\ref{difference}),
\begin{eqnarray}\label{cb}
 \frac{R_{S}}{2L}-\frac{R_{S}}{2L_{1}}=\frac{v^{2}}{c^{2}} \rightarrow\frac{v_{k}^{2}}{2c^{2}}-\frac{v^{2}}{c^{2}}\nonumber\\
                      =\frac{R_{S}}{L}[1/2-\eta^{2}]
                      \end{eqnarray}
                     where $P_{orb}\propto L^{3/2}$, then
 \begin{eqnarray}
   L_{1}=\frac{L}{1-2\eta^{2}}\Rightarrow\frac{P_{orb1}}{P_{orb}}=[1-2\eta^{2}]^{-3/2}
\end{eqnarray}
where $L_{1}$ is the new distance after the explosion and $P_{orb1}$
is the new orbital period corresponding to $L_{1}$. Fig. 2 demonstrates that $\rm
P_{orb1}$ will be evolved and  kicked up to relatively long orbital period.
(i.e.  PSR J0900-31, $\rm
P_{orb}$ = 18.7 d, e = 1.03$\times10^{-5}$ and PSR J1600-30,  $\rm P_{orb}$ = 14.35
d, and e = 1.74$\times10^{-4}$).
The above described model could actually be used for wide and circular binary
MSPs, since the AIC process in a close binary will not induce a sizeable kick velocity
to the thus formed NS$^{[17-19]}$. In our binary MSPs, we have taken the latest observational
results from the ATNF pulsar catalog$^{[20]}$, and it has been found that $\sim 67\%$  (73 out of 323) of all MSPs could be formed through the AIC process.

\begin{figure}
\includegraphics[angle=0,width=9.0cm]{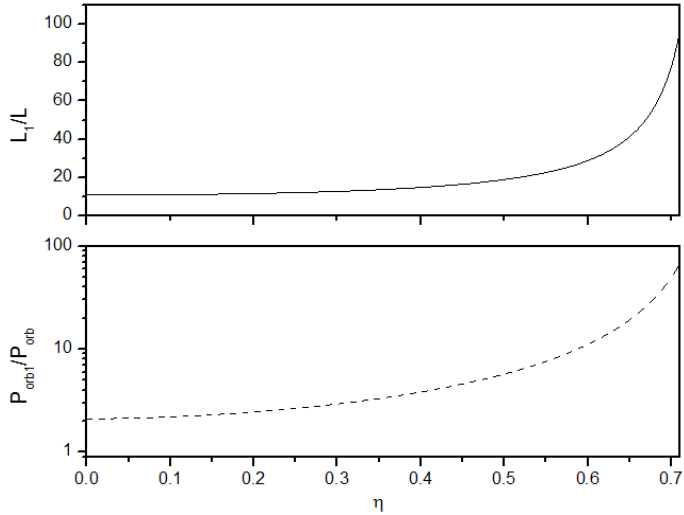}
\caption{The $\rm P_{orb1}/P_{orb}$ and $\rm L_{1}/L $ as a function
of $\eta$.}
\label{PoverP1}
\end{figure}

%It is noteworthy to mention here that the globular clusters may provide unique channels for the formation and history of accreting binaries than in the Galactic
%field. As such, the long orbital period
%binaries can originate through a tidal capture process in globular
%clusters $^{[10]}$, and there is a possibility of formation of such wide
%binaries through exchange or merger interactions$^{[1-3]}$.

\section{Summary and Conclusions}

%AIC imparts a radial kick of velocity v to the companion
%star; the corresponding increase in orbital energy propels the binary (if it
%survives) into a longer-period orbit. As such, that kick imparts a substantial
%eccentricity to the resulting orbit, as can be appreciated simply by noting
%that the point in the companion orbit at which it receives this kick must
%be common to both the pre-kick orbit and the post-kick orbit.

The dynamical instability occurs in binary systems through the AIC, is driven by dissipative processes such as kicks, mass loose and shock wave. These effects may have acquired enough energy  to the companion star and then can account for the differences in their distributions. Consequently, the AIC decreases
the mass of the NS and increases the orbital period leading to
orbit circularization. It turns out that the contribution of AIC process which potentially presents in the binary pulsars,
is about $\sim 67\%$ after the final substantial orbital eccentricities. %of all MSPs.
In addition, in some cases the kick is
sufficient to  disrupt the system, producing a population
of isolated MSPs.

%In order for this instability to be able to grow sufficiently, the delay
%between the initial formation of the proto-neutron star and the initiation of
%a "successful" explosion has to be 
%> 500ms (i.e. 100s of dynamical times),
%consistent with the most promising models of Fe core collapse to date. On the
%other hand, in the case of an e-capture supernova (§ 7.2), where the binding
%energy of the inner part of the ejecta is very small, the explosion is expected
%to occur with a much shorter delay (e.g., Kitaura et al. 2006), suggesting that
%this will produce at best a moderate kick. Since e-capture supernovae are
%more likely to occur in binary systems, this could explain why neutron stars
%in close binary sometimes appear to have received a much smaller kick than
%the majority of their single counterparts (Podsiadlowski et al. 2004).

\section*{Acknowledgements}
A. Taani would like to thank the Deanship of Scientific Research at Al-Balqa Applied university for funding his participation to this conference.

\end{document}